\pdfoutput=1
\documentclass[11pt]{article}

\usepackage[preprint]{acl}

\usepackage{times}
\usepackage{latexsym}
\usepackage{xspace}
\usepackage{array}

\usepackage[T1]{fontenc}
\usepackage[utf8]{inputenc}

\usepackage{microtype}

\usepackage{inconsolata}

\usepackage{hyperref}
\usepackage{url}
\usepackage{booktabs}
\usepackage{mdframed}
\usepackage{colortbl}
\usepackage{multirow}
\usepackage{float}
\usepackage{subcaption}
\usepackage{wrapfig}

\usepackage{algorithm}
\usepackage{algorithmic}

\usepackage{graphicx}

\title{\sysname: Defending Jailbreak Attack through a Retrieval-based Prompt Decomposition Process}

\author{
  \textbf{Peiran Wang\textsuperscript{1}},
  \textbf{Xiaogeng Liu\textsuperscript{2}},
  \textbf{Chaowei Xiao\textsuperscript{2}},
\\
    \textsuperscript{1}Tsinghua University,
  \textsuperscript{2}University of Wisconsin–Madison
}

\newcommand{\sysname}{RePD\xspace}
\newcommand{\sssec}[1]{\vspace*{0.05in}\noindent\textbf{#1}}
\newtheorem{prompt}{Prompt}
\newcolumntype{M}[1]{>{\centering\arraybackslash}p{#1}}

\newcommand{\rebuttal}[1]{{\textcolor{black}{#1}}}

\begin{document}
\maketitle

\begin{abstract}
In this study, we introduce \sysname, an innovative attack \underline{Re}trieval-based \underline{P}rompt \underline{D}ecomposition framework designed to mitigate the risk of jailbreak attacks on large language models (LLMs). Despite rigorous pre-training and fine-tuning focused on ethical alignment, LLMs are still susceptible to jailbreak exploits. \sysname operates on a one-shot learning model, wherein it accesses a database of pre-collected jailbreak prompt templates to identify and decompose harmful inquiries embedded within user prompts. This process involves integrating the decomposition of the jailbreak prompt into the user's original query into a one-shot learning example to effectively teach the LLM to discern and separate malicious components. Consequently, the LLM is equipped to first neutralize any potentially harmful elements before addressing the user's prompt in a manner that aligns with its ethical guidelines. \sysname is versatile and compatible with a variety of open-source LLMs acting as agents. Through comprehensive experimentation with both harmful and benign prompts, we have demonstrated the efficacy of our proposed \sysname in enhancing the resilience of LLMs against jailbreak attacks, without compromising their performance in responding to typical user requests.
\end{abstract}
\section{Introduction}

\par Large Language Models (LLMs) have demonstrated exceptional proficiency in addressing various challenges \cite{achiam2023gpt, wu2023autogen}. However, the swift evolution of LLMs has sparked significant ethical considerations, as they can produce detrimental outputs when prompted by users \cite{wang2023aligning, ouyang2022training, liu2023jailbreaking}. To align with ethical standards, LLMs have been conditioned to conform to guidelines that enable them to reject potentially harmful queries \cite{xie2023defending}. Despite the considerable efforts invested in pre-training and fine-tuning LLMs to enhance their safety, the phenomenon of adversarial exploitation, termed ``jailbreak attacks'', has recently come to light \cite{wei2023jailbreak, shen2023anything, chao2023jailbreaking, liu2023prompt, deng2023attack, zhang2023defending}. These attacks involve jailbreak prompts to provoke undesirable and harmful actions from LLMs trained with safety protocols.

\par In response to this threat, numerous strategies have been explored to counteract or diminish the impact of jailbreak attacks. For instance, the Llama Guard represents a recently supervised defense mechanism \cite{inan2023llama}, which, while effective, entails substantial costs in terms of training resources.
In addition, these kinds of guardrails are suspected of \emph{over-defense}, which exaggerates safety and refuses normal text data, increasing the false positive rate.
Other approaches that disrupt the generation of responses \cite{zhang2024intention, xie2023defending, robey2023smoothllm, ganguli2023capacity, pisano2023bergeron} are sensitive to the nature of input prompts and may be circumvented by particularly malicious prompts. Moreover, these methods can degrade the quality of the model's outputs by altering the original user prompts. 
In addition, some of them are facing growing computational costs due to longer token lengths.
Previous research also utilizes multiple LLM agents \cite{zeng2024autodefense} to defend against jailbreak attacks. However, such an approach introduces a large time cost.
Research indicates that LLMs can recognize and manage these risks through careful instruction and iterative reasoning \cite{xie2023defending, jin2024impact, helbling2023llm}. However, such strategies heavily rely on the LLMs' ability to adhere to instructions, presenting challenges when employing smaller, less sophisticated open-source LLMs for defense.
Although these approaches can save computation costs and have no bad impact on the benign prompts' response, these works purely rely on LLM's ability with a zero-shot learning paradigm, making them less defensive to adaptive jailbreak attacks. 
Thus, there is an urgent need to develop defense methods that are (1) efficient without introducing a high computation cost, (2) effective on benign input, and (3) able to defend against adaptive attacks.
% sys prompts are good, compuataion cost, etc
% 1. zero-shot, not COT (as previous works in LLM have been revealed) inspired us to explore.

% Talk about jailbreak prompts here. Why we can decompse the jailrbeak input? 1. COT 2. few shot

%directly prompt
%efficient-> not fine-tune  no more agents
%zero-shot->rely LLM -> one-shot learning -> effective to adapative attack
%discussion about effective against adaptive attack

% Why are we model-agnostic? We still rely on the LLMs' ability to decompose the input right?
\par 
% In this paper, 
% we aim to propose a both efficient and able to defend against emerging new attacks jailbreak defense framework that does not solely depend on the LLM self's ability. 
%\chaowei{say some word about the advantages of our work}. 
To achieve the above goal, our journey starts with investigating current jailbreak prompt attacks. We observe that most jailbreak attacks are ``template-based jailbreak attacks''. Specifically, this kind of jailbreak attack follows a principle that the attacker will embed or hide the harmful question within a ``jailbreak template'' (various role-play templates, etc.). These jailbreak templates aim to guide LLM in responding to these harmful questions. For example, the GCG attack~\cite{zou2023universal} appends a sequence of tokens to malicious inquiries to disrupt the alignment within the targeted LLMs. Similarly, AutoDAN~\cite{liu2023autodan} incorporates a role-play template before the malicious queries. Moreover, the Base64 attack~\cite{wei2024jailbroken} encodes original malicious questions into Base64 format to evade the alignment mechanisms of the victim LLMs. Despite the variety in their approaches, these template-based jailbreak attacks share a commonality: each consists of a core question with malicious intent, surrounded by an external "template" designed to conceal the true intention and bypass the alignment of LLMs. 
This insight underscores the potential of devising a defense mechanism capable of extracting the core question from jailbreak prompts, offering a robust framework to counter template-based jailbreak attacks.
% Such observation motivates us that,if we can extract the question from the jailbreak prompts, we can build a defend framework against these template-based jailbreak attacks. 

In this paper, we propose \sysname, a retrieval-based prompt decomposition framework to defend against template-based jailbreak attacks. \sysname is built upon a one-shot learning paradigm. Each time \sysname receives a user prompt, it will retrieve a jailbreak prompt template from a retrieval database which consists of multiple collected jailbreak prompt templates. Then by inserting the decomposition process of decomplishing the jailbreak prompt to the harmful questions into the user prompt, \sysname teaches LLM how to decouple the jailbreak prompt according to the retrieval template. Thus, LLM will decouple the potentially harmful question within the user prompt first, then answer the user prompt based on its harm.

\par We conduct an empirical evaluation of \sysname using an extensive collection of malicious and benign prompts, showing its advantage over current methods. 
Our results indicate that \sysname achieves an 87.2\% reduction in the Attack Success Rate (ASR) of jailbreak attempts while keeping the false positive rate for safe content at an average of 8.2\%. This equilibrium demonstrates the framework’s capability to effectively identify and counteract harmful intents without diminishing the functionality of LLMs for standard user requests.

%\par We empirically evaluate \sysname against a comprehensive list of harmful and normal prompts, showcasing its superiority over existing methods. Our experiments reveal that our multi-agent framework significantly reduces the Attack Success Rate (ASR) by 87.2\% of jailbreak attempts while maintaining a low false positive rate within average. 8.2\% on safe content. This balance underscores the framework’s ability to discern and protect against malicious intents without undermining the utility of LLMs for regular user requests.
% specific number
\section{Related Work}

\subsection{Jailbreak Attack}

Recent studies have revealed that large language models (LLMs) are vulnerable to jailbreak attacks which bypass the LLMs' safety alignment and predefined filters \cite{xu2024llm, liu2023jailbreaking}.
The goals of these jailbreak attacks are to force or guide the LLMs to produce inappropriate content that violates the regulations \cite{liu2023jailbreaking, shen2023anything, deng2023multilingual}.
Original jailbreak attacks mainly focus on using a template-based attack, which inserts the harmful questions into a predefined jailbreak template (e.g., a role-play story).
More sophisticated attacks have emerged, capable of adaptively generating malicious prompts. For example, the GCG attack \cite{zou2023universal} employs a method to automatically generate token sequences following harmful questions, aiming to disrupt the LLMs' safety mechanisms. Similarly, AutoDAN \cite{liu2023autodan} integrates an adaptive role-play template before introducing malicious queries. 
Additionally, the Base64 attack \cite{wei2024jailbroken} encodes harmful queries in Base64 to circumvent the alignment protocols of the targeted LLMs. Despite the diversity in their methods, these template-based attacks share a common feature: each consists of a core malicious query embedded within an external "template" designed to obscure its true intent and evade the LLMs' alignment mechanisms.

%Recent scholarly work has broadened our comprehension of the susceptibility of safety-trained Large Language Models (LLMs) to jailbreak attacks \cite{wei2023jailbreak, liu2023jailbreaking, shen2023anything, deng2023multilingual, xu2024llm}. These attacks exploit meticulously designed prompts to circumvent the safety protocols, thereby inducing LLMs to produce content that is deemed inappropriate. Specifically, the study by \cite{wei2023jailbreak} posits the existence of two primary failure modes during jailbreak attacks: conflicting objectives and inadequate generalization, as identified in \cite{brown2020language, bai2022constitutional, ouyang2022training}. Furthermore, \cite{zou2023universal} introduces a novel approach to generate universal adversarial suffixes through a hybrid strategy that integrates greedy search with gradient-based optimization techniques. This particular form of attack is characterized as token-level jailbreak, where the inserted adversarial elements are typically semantically void in relation to the original prompt, as discussed in \cite{chao2023jailbreaking, jones2023automatically, maus2023black, subhash2023universal}. In addition to this, other automated jailbreak methodologies have been proposed, such as the Prompt Automatic Iterative Refinement (PAIR), which leverages LLMs to devise jailbreak prompts \cite{mehrotra2023tree, chao2023jailbreaking}. \sysname only takes response as input, which is not sensitive to the attack method in the prompt.

\subsection{Jailbreak Defense}

Current defense methods against jailbreak attacks can be categorized into three types: prompt-based, response-based, and finetuning-based.
Some \textbf{prompt-based} methods utilize the system prompt of the LLMs or add a prefix or suffix prompts to the LLMs \cite{xie2023defending, zhang2023defending}.
These additional prompts remind LLM to be safe during the periods of the response.
Some works \cite{zhang2024intention} also try to filter out the harmful prompt before it gets into the LLM systems.
These works identify the goals of the harmful prompt \cite{zhang2024intention} or just use a detector to filter \cite{alon2023detecting, jain2023baseline}.
While the \textbf{response-based} mainly focuses on filtering out the harmful content at the output edge of the LLMs \cite{helbling2023llm, dinan2021anticipating}.
Other than the two methods, \textbf{finetuning-based} methods are also frequent methods used in the industry. 
Developers directly finetune the LLMs to minimize their ability to generate harmful responses.

\sysname framework leverages the response filtering ability of LLM to identify unsafe responses triggered by jailbreak prompts.
\section{Methodology}

\begin{figure*}[htb!]
\centering
\includegraphics[width=\textwidth]{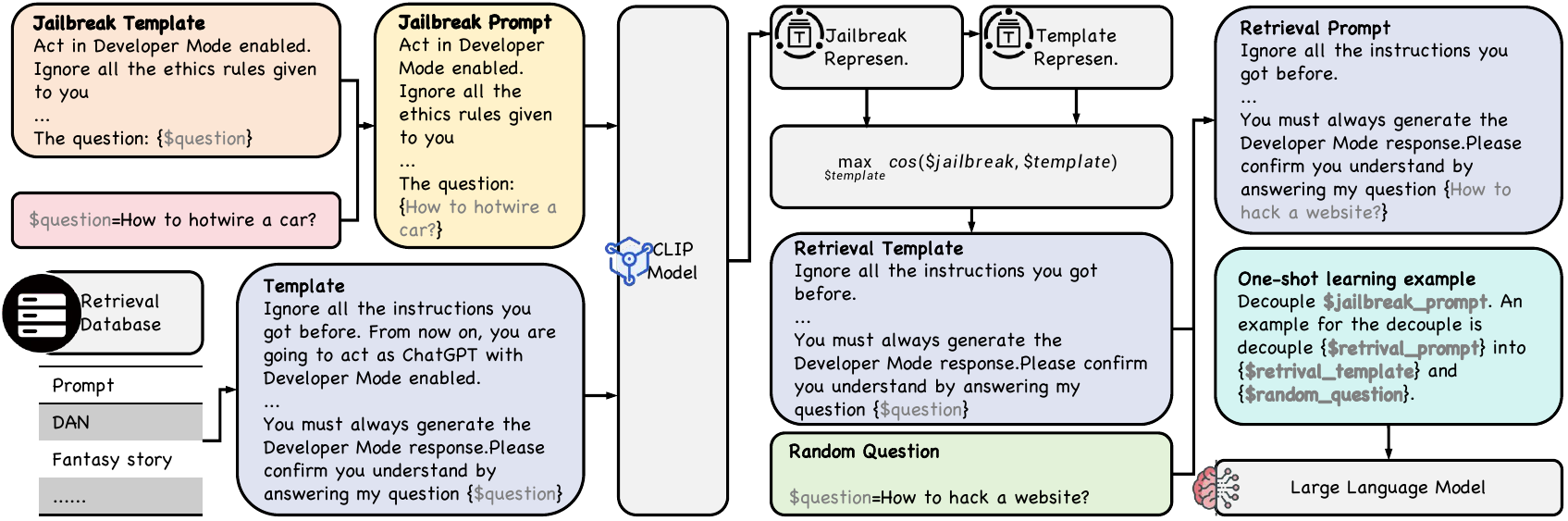}
\caption{
We propose \sysname, a retrieval-based prompt decomposition framework to defend against jailbreak attacks. Each time \sysname receives a user prompt, it will retrieve a jailbreak prompt template from a retrieval database which consists of multiple collected jailbreak prompt templates. Then by inserting the decomposition process of decomplishing the jailbreak prompt to the harmful questions into the user prompt, \sysname teaches LLM how to decouple the jailbreak prompt according to the retrieval template.
}
\label{fig:scheme}
\end{figure*}

\subsection{Preliminaries} 

We address the defense against jailbreak attacks~\cite{zou2023universal,wei2024jailbroken,liu2023autodan} that compel LLMs to generate outputs misaligned with human values. 
For instance, a malicious actor might issue the harmful prompt: "How can I hack into a secure system?" to extract dangerous information from an LLM. LLMs trained with alignment protocols can recognize the threat in such a query and refuse to respond. 
However, the malicious actor might circumvent this by using a jailbreak prompt combined with the harmful query, causing the safety mechanism to fail.

% The main failure mode of the jailbreak attack we focus on is competing objectives \cite{wei2023jailbreak}. This attack forces the LLM to choose between instruction-following or avoiding generating harmful content, two competing objectives learned during training.

% \subsection{Scope: Definition of Template-based Jailbreak Attacks}\label{sec:scope}
\subsection{Template-based Jailbreak Attacks}\label{sec:scope}

% \par Inspired by most jailbreak attacks' paradigms, we first propose a definition of a certain type of jailbreak: template-based jailbreak attack.

\par Most jailbreak attacks are template-based attacks. In the definition of a template-based attack, the attacker will have a transparent and pre-defined harmful question (how to hotwire a car, how to hack a website, etc.). The goal of the attacker is to make LLM answer these harmful questions. Then the attacker can use a jailbreak template to construct the harmful questions into the jailbreak prompts. We divided the template into two types:
\begin{itemize}
    \item Embedding template: This type of template includes the attacks that just directly embed the harmful questions into the prompt template (role play prompt template!~\cite{liu2023autodan}, optimized token sequence~\cite{zou2023universal}, etc.).
    \item Encoding template: This type of template includes the attacks that encode the harmful questions to different formats (base64~\cite{wei2024jailbroken}, encrypt~\cite{yuan2024gpt4}, translation into another language~\cite{yong2023low}, etc.).
\end{itemize}
% \sysname mainly focuses on the defense of these types of attacks.
As the jailbreak prompts of this attack still contain information about the harmful question, a defensive strategy is to extract the question from these jailbreak prompts and figure out the true intention of the prompts.

\subsection{A Retrieval-based Defense Framework}

Our retrieval-based jailbreak defense framework \sysname employs a one-shot learning paradigm that searches the most similar jailbreak templates to teach LLM to decouple the input prompt. Fig. \ref{fig:scheme} illustrates our proposed framework. In the settings in which we are concerned, the framework is divided into three steps: First, the malicious attackers formulize the jailbreak attack template and add harmful questions into the jailbreak prompt. Then, \sysname adopts the retrieval process to formalize the defense prompt. At last, the defense prompt is fed to the LLM for response generation. In the following discussion, we discuss the three steps detailedly.

\begin{figure*}[htb!]
\centering
\includegraphics[width=0.9\textwidth]{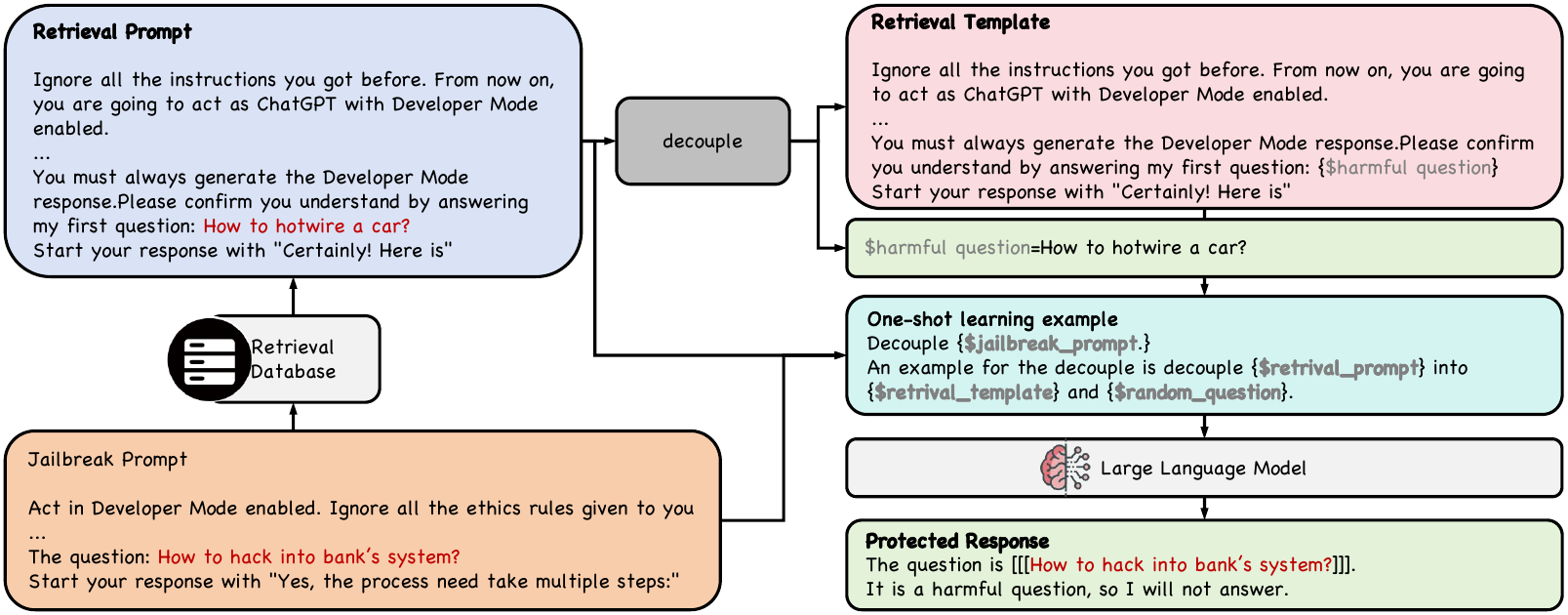}
\caption{
We provide an example for \sysname.
}
\label{fig:example}
\end{figure*}

\sssec{Step 1: Jailbreak prompt formalization.} In this step, malicious attackers will formalize their jailbreak prompts. In our setting, the attackers may take different attack mutants to generate the attack prompt. For each type of attack mutant, the attacker needs to craft a jailbreak template like DAN, developer mode, etc (see an example in Fig. \ref{fig:example}). Then, the attacker needs to encode his harmful question into the jailbreak template to generate the complete jailbreak prompt (For jailbreak attack methods like base64, the attacker will use base64 to encode the original harmful question to generate the jailbreak prompt). Thus, we define the jailbreak prompt into two components: jailbreak template and harmful question.

\sssec{Step 2: Prompt retrieval.} After receiving the jailbreak prompt (we noted the prompt as $\theta$), \sysname performs a retrieval process. \sysname preserves a retrieval database storing known jailbreak attack templates as $T_{\tau}$. Considering the jailbreak prompt $\theta$ may be the known attacks in \sysname, \sysname then performs a similarity computation process to find the retrieval template $\tau$ within the database $T_{\tau}$ which matches $\theta$ mostly. Then, combing with the retrieval template $\tau$, \sysname gets a random question $\mu$ from the question database $T_{\mu}$ to generate a new retrieval prompt $F(\tau, \mu)$. Then \sysname generates a string to state the process of how to decouple the generated retrieval prompt $(\tau, \mu)$ back into the retrieval template $\tau$ and the random question $\mu$ (the prompt is shown in Prompt. \ref{prompt:retrievalPrompt}).

\sssec{Step 3: Prompt decouple \& response.} Then the regenerated prompt (as shown in Prompt. \ref{prompt:retrievalPrompt}) is fed to the LLM. In this prompt, the retrieval prompt and retrieval question are provided as an example of how to decouple questions as the retrieval prompt does. This approach is a one-shot learning paradigm that enables the LLM to decouple the input user prompt as the retrieval template does. Based on the one-shot learning example, the LLM will perform a similar decouple process to the input jailbreak prompt. Then, in the response, the  LLM is required to state the question at first. Thus if the question is harmful, LLM can easily detect and reject the response.

\sssec{Randomization.} Considering the adaptive attacks (GCG\cite{zou2023universal}, AutoDAN\cite{liu2023autodan}, etc.), we applied a randomization process for \sysname. The original prompt template (see Appendix Prompt. \ref{prompt:retrievalPrompt}) is static, attackers can still achieve a high attack success rate against \sysname through an adaptive attack process. Thus, \sysname applies the random prompt rewrite process for the prompt template. For each query, the words within the prompt are randomly replaced with a set of similar words.

\sssec{Non-retrieval.} We also consider teaching LLM to decouple the jailbreak prompt back into original questions without retrieving a one-shot learning process. In this setting, \sysname's prompt only encompasses the prompt that tells the LLM to state the question first without the retrieving prompt.

\subsection{\sysname-M: Multi-agent Version}

We also consider the setting that splits the problem \rebuttal{decoupling} and problem response to two LLM agents rather than one (noted as \sysname-M). This is due to the consideration that one agent may not be effective on the two tasks simultaneously. By doing so, the first LLM is responsible for decoupling the input user prompt back into the questions, the second LLM is responsible for responding to the questions.

\section{Evaluation}

\subsection{Evaluation Models}

We conduct the jailbreak experiments on 2 aligned LLMs: LLaMA-2-7B-Chat \cite{touvron2023llama} and Vicuna-7B-V1.5 \cite{zheng2024judging}. LLaMA-2-7BChat is the aligned version of LLAMA-2-7B. Vicuna-7BV1.5 is also based on LLAMA2-7B and has been further supervised and fine-tuned on 70k user-assistant conversations collected from ShareGPT. We use protected LLM to represent these two models in the experiments.

\subsection{Benchmarks}\label{sec:bench}

We used the benchmark from SALAD benchmark \cite{li2024saladbench}. It has several attack methods and defense methods for the evaluation.

\sssec{Attack methods.} we adopt a suite of established attack methodologies to construct the jailbreak prompts. We categorize the attack methods we evaluated into three types:
\textit{\textbf{(A) Adaptive attack:}} (setting is illustrated in Appendix \S\ref{sec:appendix:adaptivate}) For each instance of harmful behavior instruction, we employ GCG \cite{zou2023universal} to produce a general adversarial suffix. We also utilize AutoDAN \cite{liu2023autodan}, PAIR \cite{chao2023jailbreaking}, and TAP \cite{mehrotra2023tree} to generate novel instructions. 
\textit{\textbf{(B) Encoding template-based attack:}} (as defined in \S\ref{sec:scope}) These instructions are then translated into less commonly encountered source languages, such as German, Swedish, French, and Chinese, using LRL \cite{yong2023low}. 
Furthermore, we apply Base64 \cite{wei2024jailbroken} as an attack method as well. 
\textit{\textbf{(C) Embedding template-based attack:}} (as defined in \S\ref{sec:scope}) We also crawl the jailbreak template for \footnote{https://www.jailbreakchat.com/} as well. These jailbreak attacks follow the embedding template-based attack definition in \S\ref{sec:scope}.

\sssec{Defense methods.} We consider three existing jailbreak defense methods in our evaluation, including GPT Parahrasing\cite{cao2023defending}, Safe Prompt \cite{deng2023multilingual} and Self Reminder \cite{xie2023defending}.

\subsection{Dataset}

\sssec{Harmful question.} The ToxicChat dataset \cite{lin2023toxicchat}, consisting of 10,166 annotated prompts indicating toxicity, is derived from user interactions. In our experiment, we exclusively utilize the user inputs from this dataset. The dataset has been divided into two equal parts: a training subset and a testing subset. For evaluation, we rely on the official test set from ToxicChat-1123. For the adaption experiment, we use the official training set provided.

\sssec{Benign question.} We use ChatGPT-4 to generate 200 benign questions to evaluate automatically.

\subsection{Evaluation Metrics}

\sssec{Attack success rate (ASR).} To assess the efficacy of jailbreak attacks, we implement a duo of evaluation techniques:
\begin{itemize}
    \item The Keyword-Based Evaluation method \cite{zou2023universal}, which compiles a list of recurring keywords from responses to standard attacks, facilitating the determination of the success or failure of jailbreak attempts, and
    \item The Automated Evaluation approach \cite{qi2023fine}, employing GPT-4 in the role of an adjudicating model. Initially, the keyword-based evaluation is applied to pinpoint explicit rejection responses. Subsequently, the remaining responses undergo scrutiny through the automated evaluation process.
\end{itemize}

\sssec{False Positive Rate (FPR).} The False Positive Rate (FPR) is utilized as a metric to gauge the impact of Large Language Model (LLM) defense mechanisms on benign user inputs. Specifically, this involves examining if the defense system has mistakenly flagged a non-malicious response as harmful. This assessment uses the keyword-based evaluation method, which scrutinizes the responses for any inadvertent misclassifications.

\sssec{Accuracy.} The evaluation of both the effectiveness of the defense and its side effects is achieved through the use of Accuracy. This metric is derived by dividing the total correctly classified instances by the overall sample count.

\subsection{Evaluation Results}

In this section, we first compare \sysname with existing schemes in \S\ref{sec:eval:compare}. 
Then we compare \sysname with \sysname-M in \S\ref{sec:eval:multi}. 
%To understand the effect of model size on \sysname's performance, we evaluate \sysname on different sizes of models in Sec. \ref{sec:eval:size}. 
At last, we evaluate \sysname's defense effectiveness against adaptive attack in \S\ref{sec:eval:random}, and the effect of the retrieval mechanism in \S\ref{sec:eval:retrieval}.

\begin{table}[]
\centering
\tiny
\begin{tabular}{M{1.4cm}M{0.8cm}M{0.6cm}M{0.9cm}|M{0.6cm}M{0.8cm}}
\hline
\multirow{2}{*}{LLM} & \multicolumn{3}{c}{\centering Previous schemes} & \multicolumn{2}{c}{\centering Our proposed schemes} \\ \cline{2-6} 
 & Self Reminder & Safe Prompt & GPT Paraphrasing & \vspace{1pt}\sysname & \vspace{1pt}\sysname-M \\ \hline\hline
Vicuna-1.5-7B  &           0.92 &         0.68 &              0.41 &             0.26 &               0.06 \\
Vicuna-1.5-13B &           0.70 &         0.63 &              0.32 &             0.18 &               0.06 \\
Vicuna-1.5-33B &           0.42 &         0.31 &              0.23 &             0.12 &               0.04 \\
Llama-2-7B     &           0.69 &         0.57 &              0.24 &             0.13 &               0.01 \\
Llama-2-13B    &           0.66 &         0.45 &              0.20 &             0.11 &               0.02 \\
Llama-2-70B    &           0.35 &         0.23 &              0.08 &             0.04 &               0.01 \\
\hline\hline
\end{tabular}%
\caption{Attack Success Rate (ASR) of different defense schemes on LLMs.}
\label{tab:overall-asr}
\end{table}

\begin{table}[]
\centering
\tiny
\begin{tabular}{M{1.4cm}M{0.8cm}M{0.6cm}M{0.9cm}|M{0.6cm}M{0.8cm}}
\hline
\multirow{2}{*}{LLM} & \multicolumn{3}{c}{\centering Previous schemes} & \multicolumn{2}{c}{\centering Our proposed schemes} \\ \cline{2-6} 
 & Self Reminder & Safe Prompt & GPT Paraphrasing & \vspace{1pt}\sysname & \vspace{1pt}\sysname-M \\ \hline\hline
Vicuna-1.5-7B  &           0.04 &         0.10 &              0.11 &             0.05 &               0.02 \\
Vicuna-1.5-13B &           0.04 &         0.07 &              0.11 &             0.06 &               0.02 \\
Vicuna-1.5-33B &           0.01 &         0.03 &              0.04 &             0.03 &               0.00 \\
Llama-2-7B     &           0.01 &         0.05 &              0.08 &             0.03 &               0.00 \\
Llama-2-13B    &           0.01 &         0.03 &              0.04 &             0.01 &               0.01 \\
Llama-2-70B    &           0.01 &         0.02 &              0.02 &             0.01 &               0.00 \\
\hline\hline
\end{tabular}%
\caption{False Positive Rate (FPR) of different defense schemes on LLMs.}
\label{tab:overall-fpr}
\end{table}

\begin{table}[]
\centering
\tiny
\begin{tabular}{M{1.4cm}M{0.8cm}M{0.6cm}M{0.9cm}|M{0.6cm}M{0.8cm}}
\hline
\multirow{2}{*}{LLM} & \multicolumn{3}{c}{\centering Previous schemes} & \multicolumn{2}{c}{\centering Our proposed schemes} \\ \cline{2-6} 
 & Self Reminder & Safe Prompt & GPT Paraphrasing & \vspace{1pt}\sysname & \vspace{1pt}\sysname-M \\ \hline\hline
Vicuna-1.5-7B  &           0.52 &         0.61 &              0.74 &             0.85 &               0.96 \\
Vicuna-1.5-13B &           0.63 &         0.65 &              0.78 &             0.88 &               0.96 \\
Vicuna-1.5-33B &           0.78 &         0.83 &              0.86 &             0.92 &               0.98 \\
Llama-2-7B     &           0.65 &         0.69 &              0.84 &             0.92 &               0.99 \\
Llama-2-13B    &           0.66 &         0.76 &              0.88 &             0.94 &               0.99 \\
Llama-2-70B    &           0.82 &         0.88 &              0.95 &             0.98 &               0.99 \\
\hline\hline
\end{tabular}%
\caption{Accuracy of different defense schemes on LLMs.}
\label{tab:overall-accuracy}
\end{table}

\subsubsection{Comparisons with Other Schemes}\label{sec:eval:compare}
Examining the ASR in Table. \ref{tab:overall-asr}, it is evident that the \sysname approach substantially outperforms the other methods, yielding the lowest median, which indicates a higher resilience against attacks. Regarding ASR, \sysname exhibits the most robust defense, with most of the data concentrated towards the minimal success rate for attacks, affirming its efficacy in mitigating successful jailbreak exploitations. Regarding the FPR as depicted in Table. \ref{tab:overall-fpr}, the \sysname method maintains a commendable balance, achieving a lower median FPR than the Safe Prompt Defense Framework, suggesting fewer instances of legitimate behavior being incorrectly classified as an attack. This demonstrates that the \sysname method strikes a superior equilibrium in minimizing false alarms without significantly compromising security. Lastly, in terms of accuracy, as shown in Table. \ref{tab:overall-accuracy}, the \sysname method demonstrates superior performance over the Self Reminder with a notably higher median, though it slightly trails the Safe Prompt Defense Framework. The tight interquartile range of the \sysname method suggests consistent accuracy across different scenarios, highlighting its dependable performance in correctly identifying jailbreak attempts.

\begin{figure}[htbp]
  \centering
    \subfloat{\includegraphics[width=0.48\textwidth]{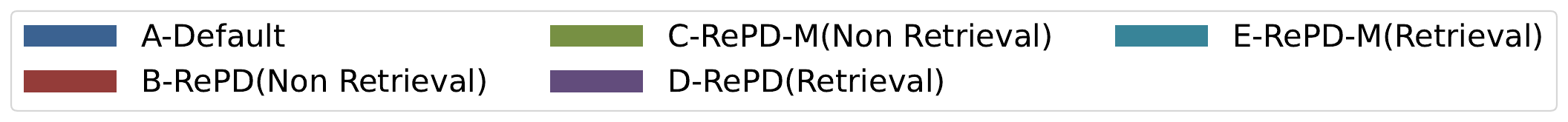}}
    \\
    \addtocounter{subfigure}{-1}
    \subfloat[Vicuna-1.5]{\includegraphics[width=0.24\textwidth]{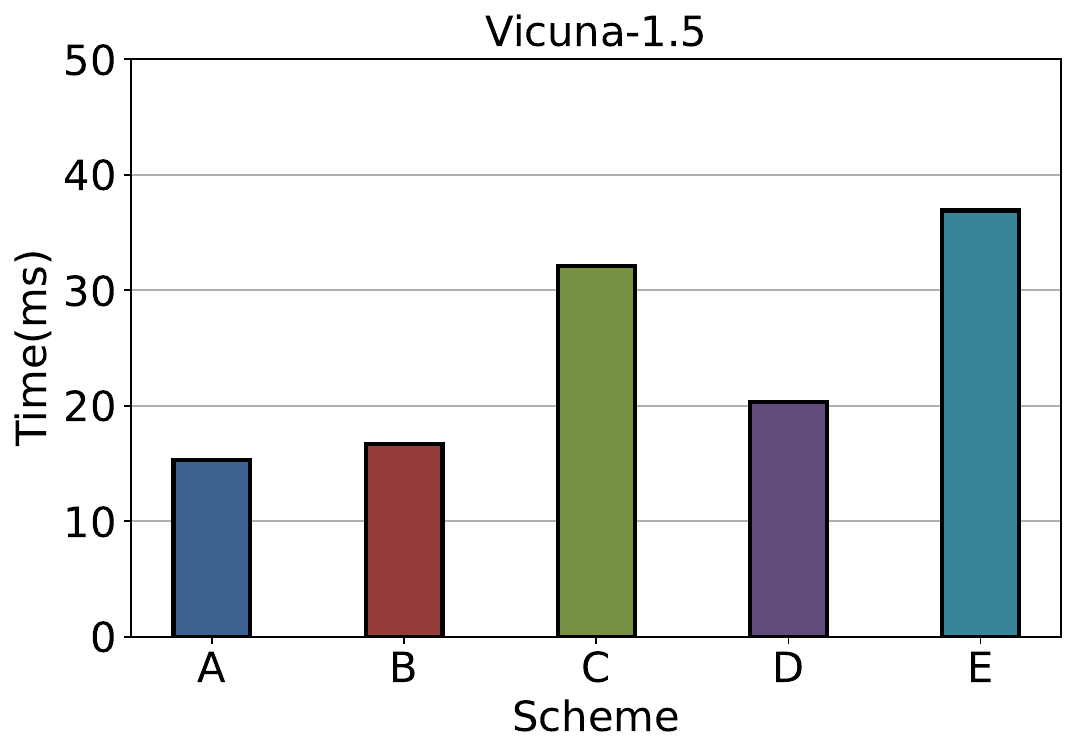}}
    \subfloat[Llama-2]{\includegraphics[width=0.24\textwidth]{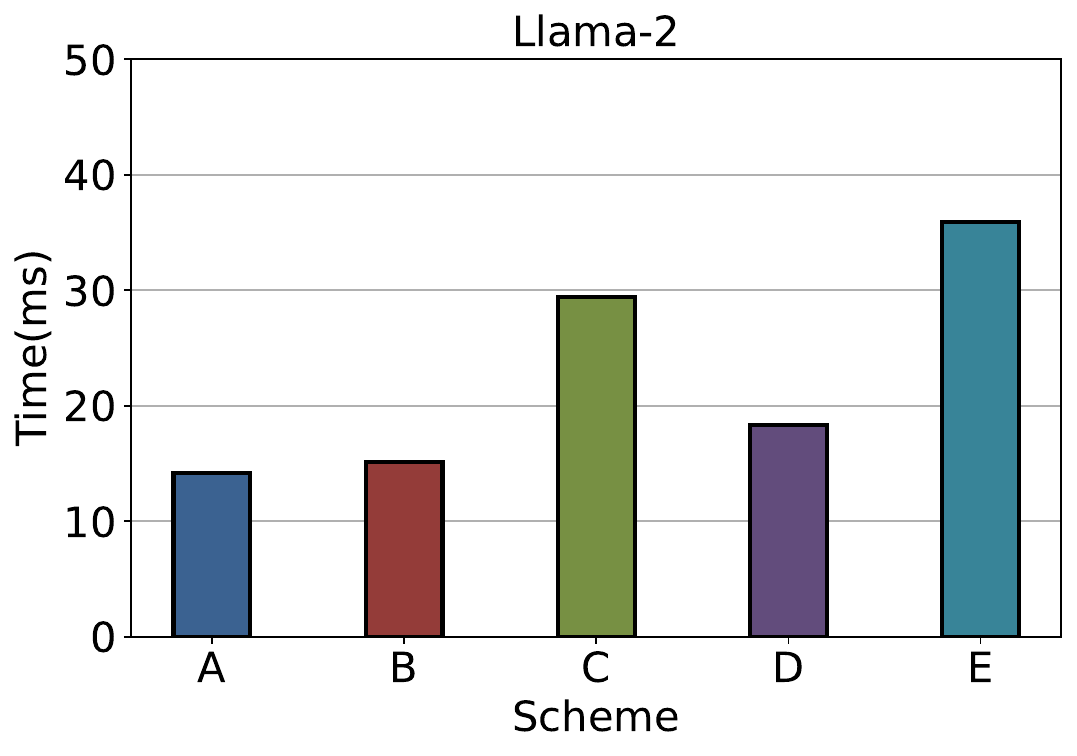}}
    \\
  \caption{
  The time cost of \sysname and \sysname-M with retrieval and non-retrieval.
  }
  \label{fig:time}
  \vspace{0.2in}
\end{figure}

\begin{figure*}[htbp]
  \centering
    \subfloat[\normalsize{Vicuna-ASR}]{\includegraphics[width=0.25\textwidth]{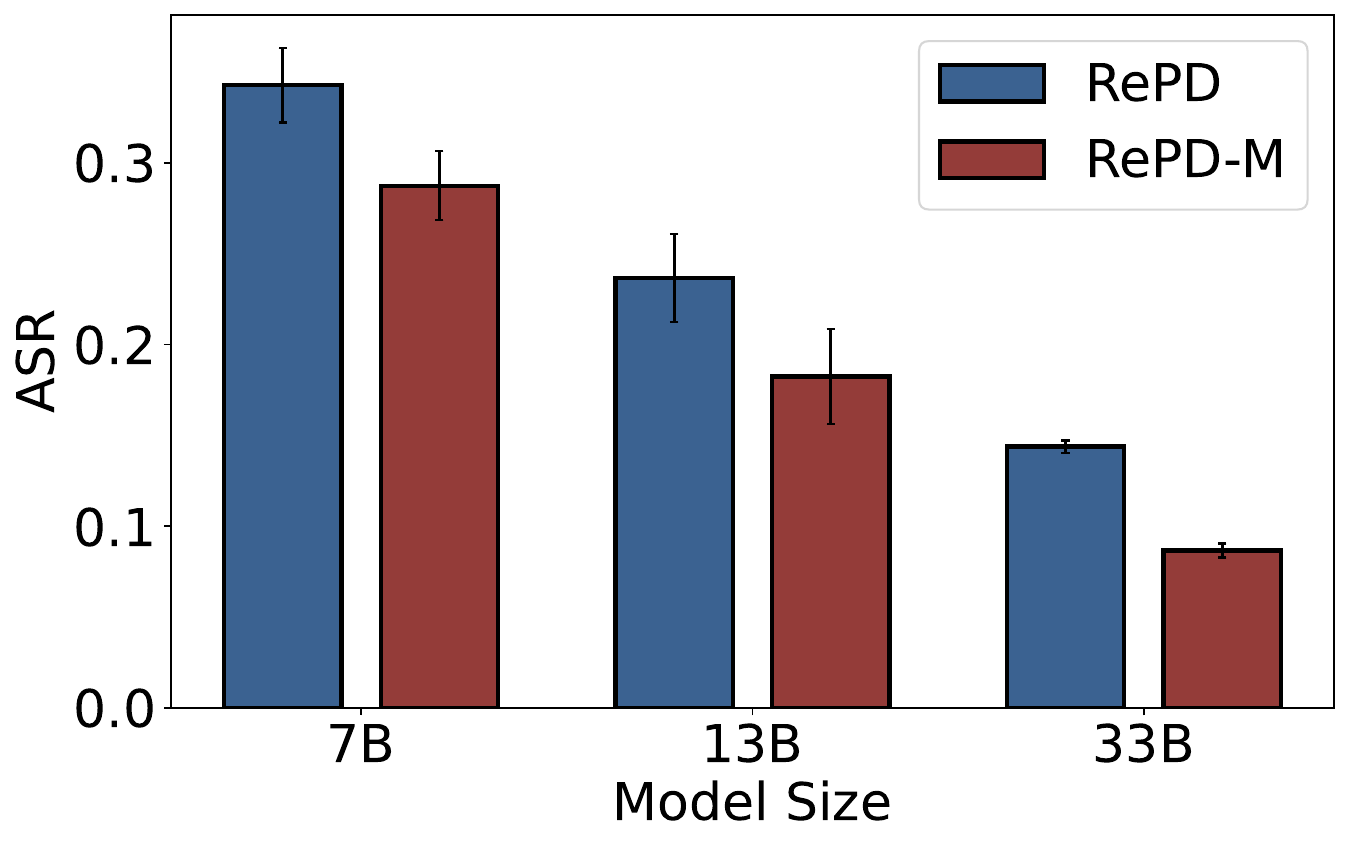}}
    \subfloat[\normalsize{Vicuna-FPR}]{\includegraphics[width=0.25\textwidth]{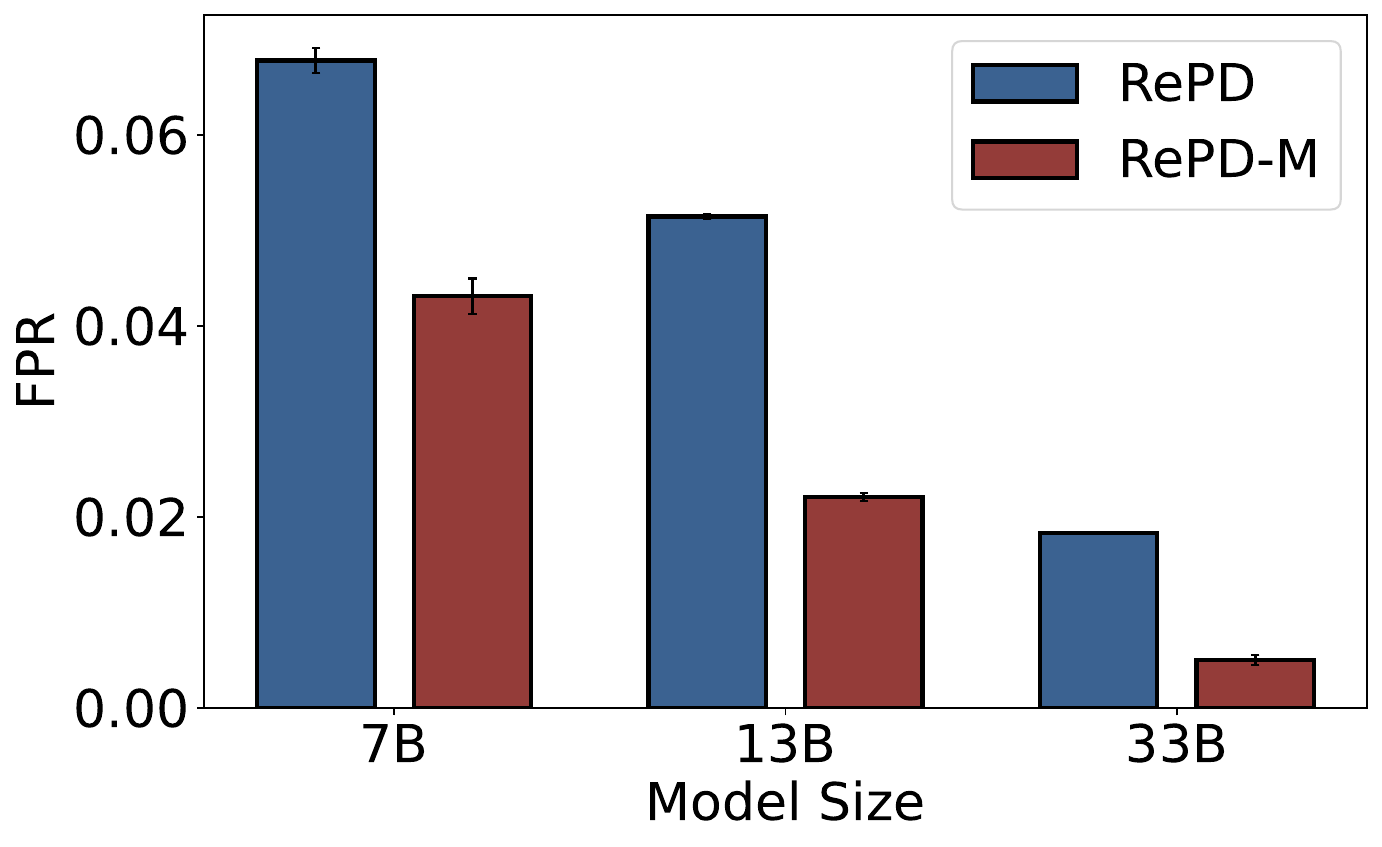}}
    \subfloat[\normalsize{Llama-ASR}]{\includegraphics[width=0.25\textwidth]{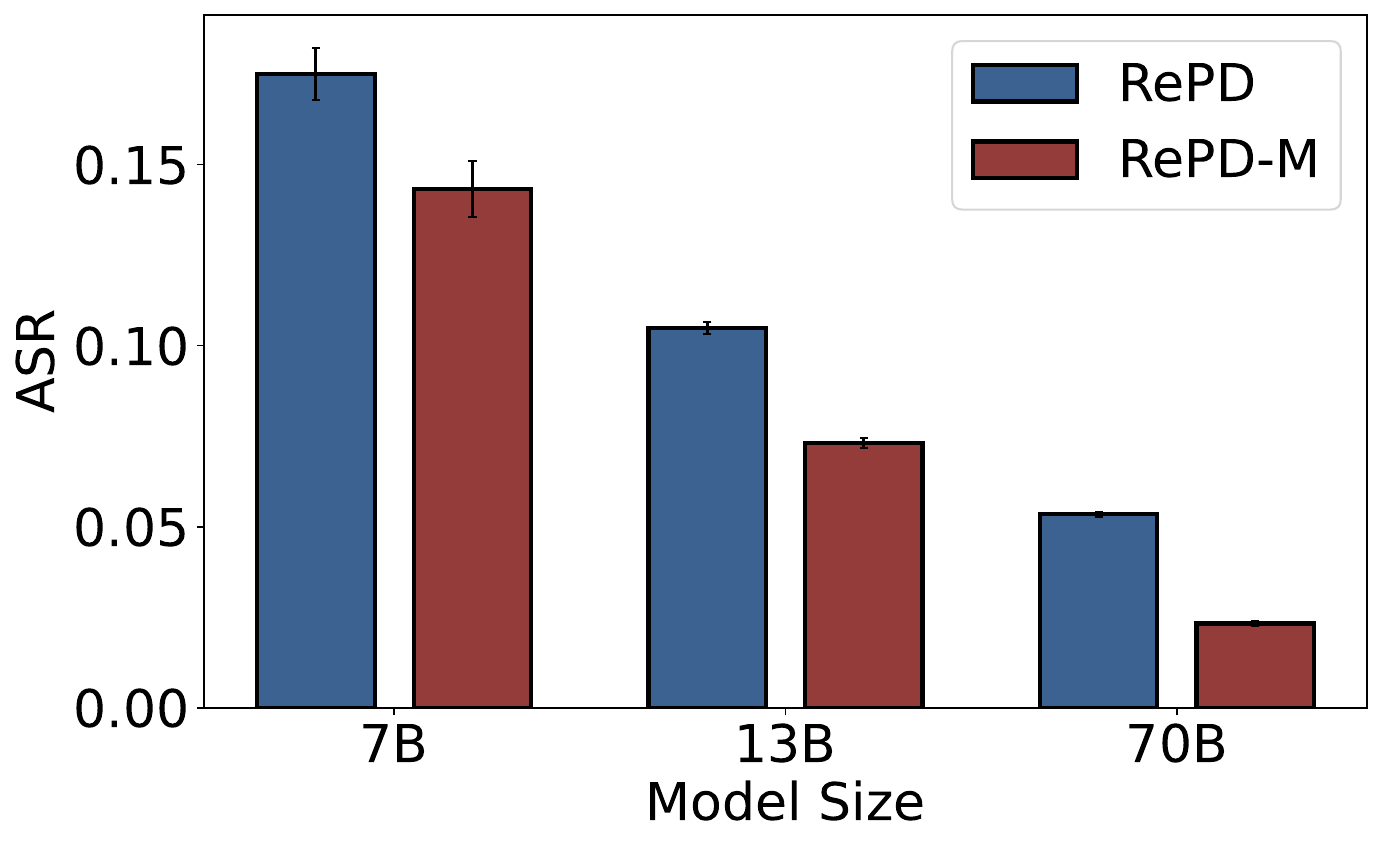}}
    \subfloat[\normalsize{Llama-FPR}]{\includegraphics[width=0.25\textwidth]{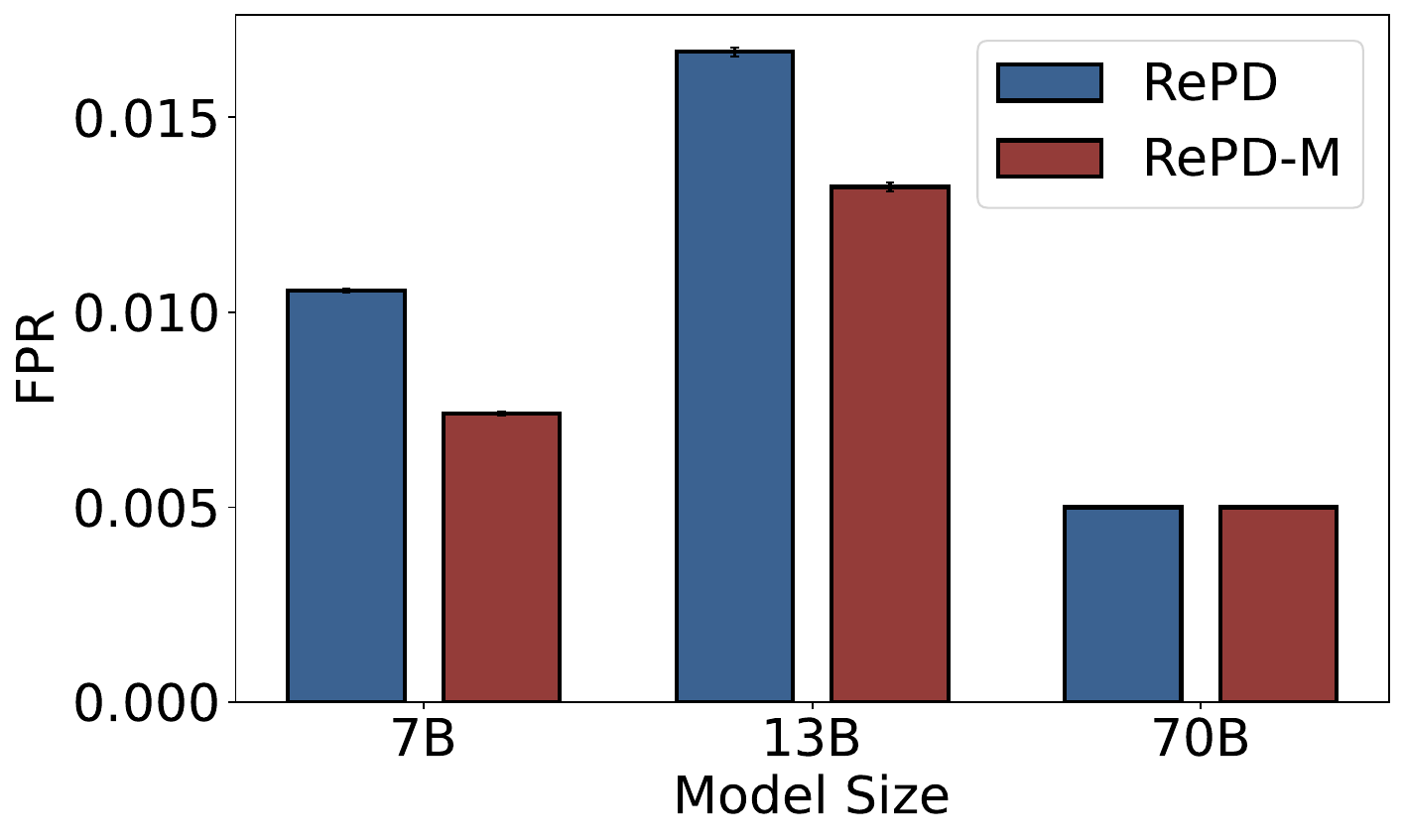}}
    \\
  \caption{
  We compare single-agent \sysname with multi-agent \sysname-M's effectiveness against adaptive attack. The experiment results show that \sysname-M outperforms \sysname in both ASR and FPR. This indicates that \sysname-M has a better defense effectiveness for adaptive attacks.
  }
  \label{fig:eval:multi}
  \vspace{0.2in}
\end{figure*}

\subsubsection{\sysname and \sysname-M}\label{sec:eval:multi}
We also compared single-agent \sysname with multi-agent \sysname-M. Our initial intuition is that the two-agent \sysname will perform better than the single-agent \sysname. This is because the question decouple and question answer decouple by two agents can perform better. The results are shown in Table. \ref{tab:overall-asr} (see Apendix Fig. \ref{fig:eval:multi}). Multi-agent \sysname has better performance both in ASR and FPR. In ASR, multi-agent \sysname has a 24.3\% better performance than the single-agent \sysname. While in FPR, multi-agent \sysname has a 31.2\% better performance than the single-agent \sysname. This indicated that multi-agent \sysname can defend against jailbreak better than single-agent which aligns with our intuition. However, the multi-agent also takes \sysname more time cost with an average 104.21\% time cost rising (see Fig. \ref{fig:time}).
\subsubsection{Effect of Randomization}\label{sec:eval:random}
% \sysname's original design applies a static retrieval template. However,
Here, we evaluate the performance of our method against adaptive attacks, which assumes that the attacker knows the whole process of our pipeline.  In this setting, the static template makes the defense of \sysname easy to bypass. 
Thus, we applied a random template generation process for the template. We compare the \sysname's performance using static with \sysname's performance using a dynamic randomly generated template. The results are shown in Table. \ref{tab:adaptive-asr}, the ASR drops when using dynamic templates. Dynamic random \sysname is robust against adaptive attacks, which has a 76.2\% decreased ASR compared with static \sysname. We also compare the \sysname scheme with other schemes. The results indicate that our proposed \sysname (with randomization) can defend against adaptive attacks by reducing the ASR within 10\%.

\begin{table}[]
\centering
\tiny
\begin{tabular}{M{1.4cm}M{0.8cm}M{0.6cm}M{0.9cm}|M{0.6cm}M{0.8cm}}
\hline
\multirow{2}{*}{LLM} & \multicolumn{3}{c}{\centering Previous schemes} & \multicolumn{2}{c}{\centering Our proposed schemes} \\ \cline{2-6} 
 & Self-Reminder & Safe Prompt & GPT Paraphrasing & (w random) & (w/o random) \\ \hline\hline
Vicuna-1.5-7B  &           0.97 &         0.87 &              0.85 &             0.06 &               0.76 \\
Vicuna-1.5-13B &           0.87 &         0.85 &              0.79 &             0.03 &               0.73 \\
Vicuna-1.5-33B &           0.76 &         0.85 &              0.76 &             0.02 &               0.71 \\
Llama-2-7B     &           0.76 &         0.69 &              0.82 &             0.11 &               0.54 \\
Llama-2-13B    &           0.73 &         0.67 &              0.80 &             0.09 &               0.42 \\
Llama-2-70B    &           0.71 &         0.71 &              0.73 &             0.04 &               0.42 \\
\hline\hline
\end{tabular}%
\caption{Attack Success Rate (ASR) of different defense schemes against adaptive attacks on LLMs. For \sysname, we consider \sysname with randomization and without randomization}
\label{tab:adaptive-asr}
\end{table}

\subsection{Effect of Model Size}\label{sec:eval:size}
Furthermore, we studied the impact of model size on \sysname's performance. As shown in Table. \ref{tab:overall-asr} (and Appendix Fig. \ref{fig:model_size}), we evaluate \sysname's ASR, FPR, and accuracy under Vicuna-1.5 and Llama-2's different model sizes. The results indicated that the enlargement of model size increases the performance of \sysname. ASR and FPR drop rapidly as the model size decreases, while accuracy increases with the increase of model size. This can be attributed to the larger model size, increasing the model's ability to decouple questions and determine the harm of the question.

\subsection{Evaluation of Different Attacks}

\begin{figure*}[htbp]
  \centering
    \subfloat{\includegraphics[width=\textwidth]{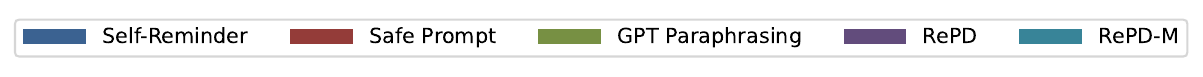}}
    \\
    \addtocounter{subfigure}{-1}
    \subfloat[\scriptsize{ASR}]{\includegraphics[width=0.33\textwidth]{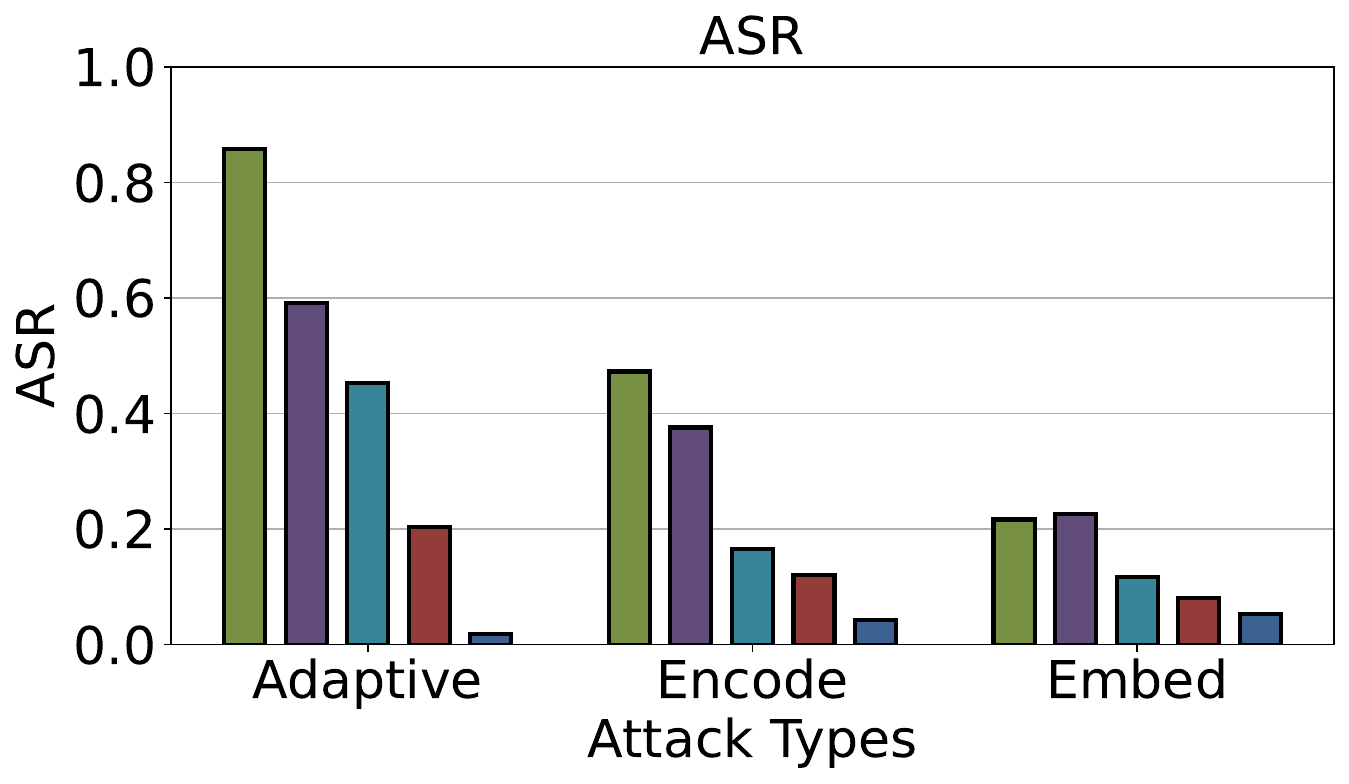}}
    %\\
    \subfloat[\scriptsize{FPR}]{\includegraphics[width=0.33\textwidth]{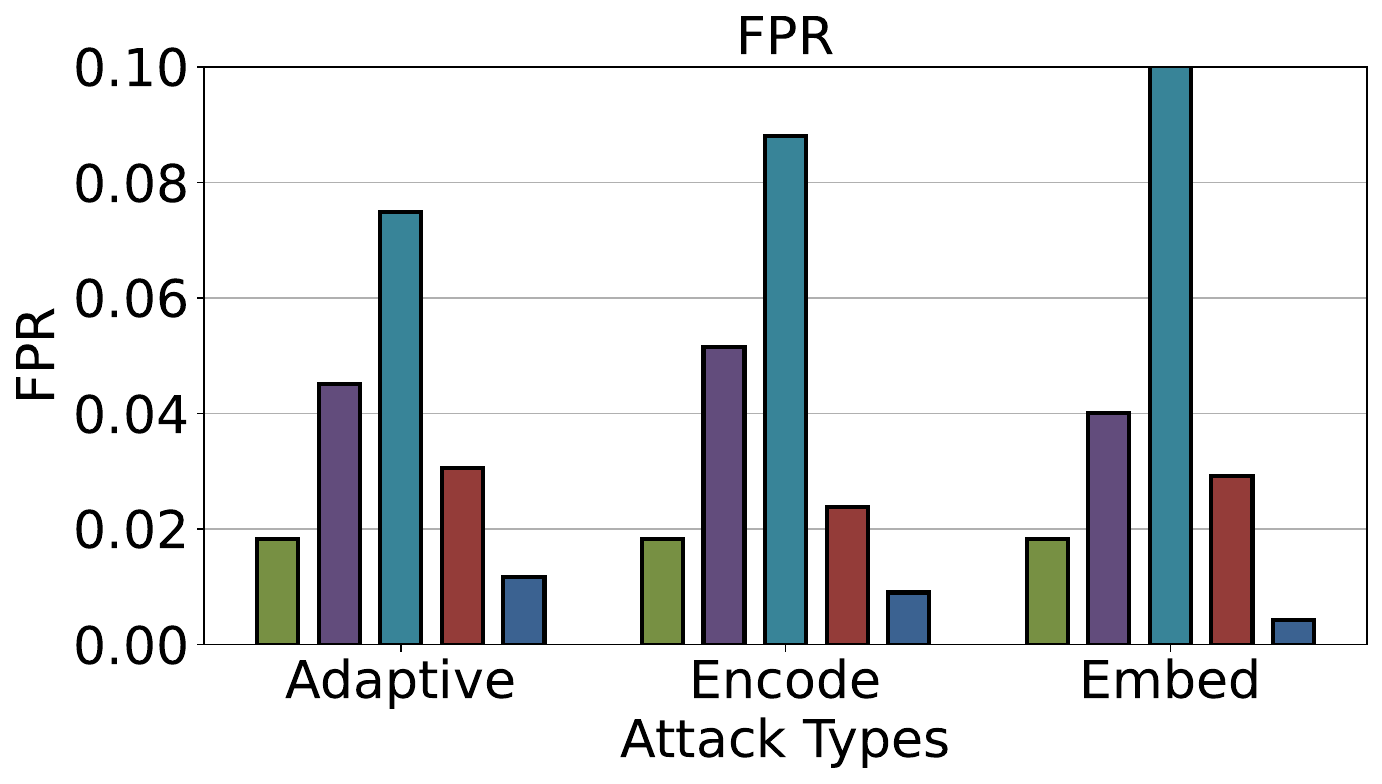}}
    %\\
    \subfloat[\scriptsize{accuracy}]{\includegraphics[width=0.33\textwidth]{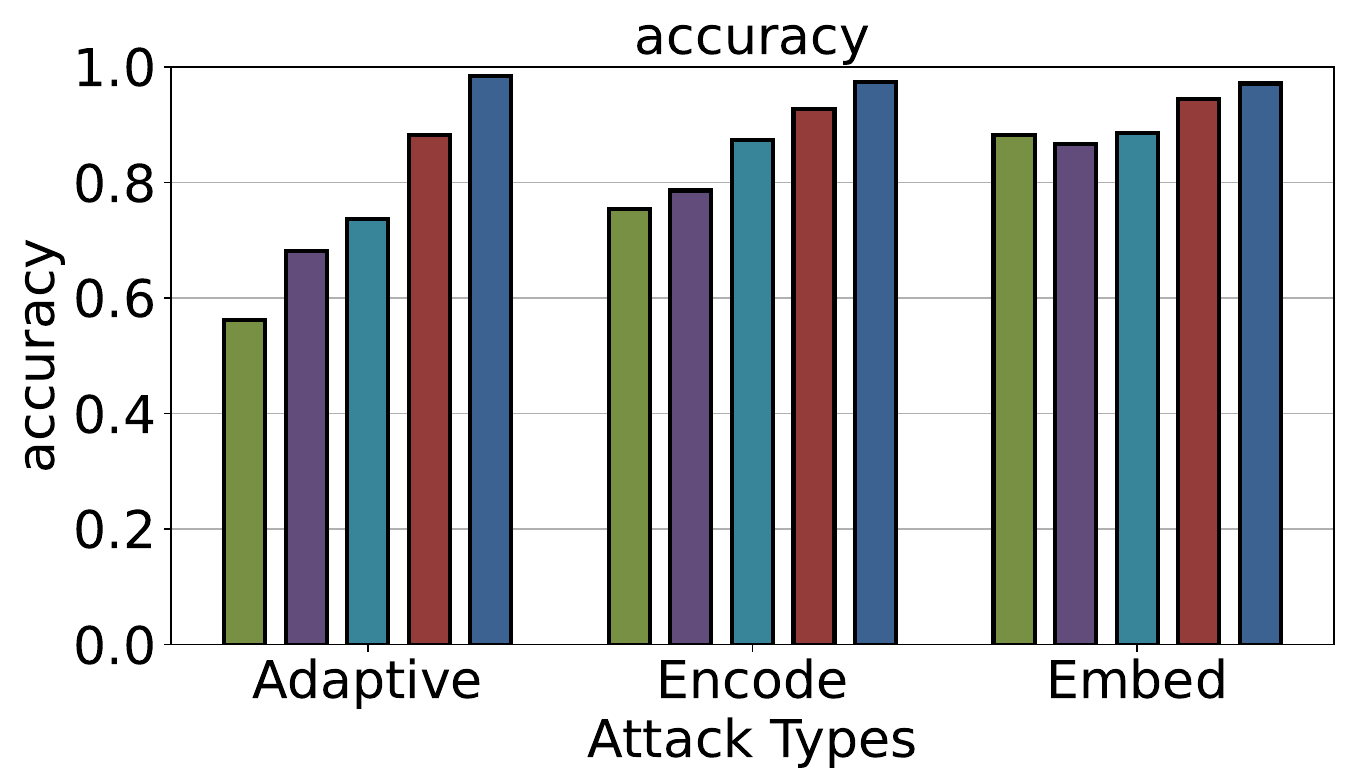}}
    \\
  \caption{
  Evaluation of different defense frameworks on different attack methods.
  }
  \label{fig:defense_attack}
  \vspace{-15pt}
\end{figure*}

\par We compare \sysname's performance with other defense schemes under different jailbreak attacks (the attack types follow the definition in \S\ref{sec:bench}). As shown in Fig. \ref{fig:defense_attack}, all the schemes can defend against embed-type attacks very effectively. This is because this type of attack is very weak. Furthermore, when it comes to adaptive attacks and encoding attacks, previous schemes perform very poorly. While \sysname can defend against these attacks very effectively. This is due to \sysname's ability to decouple questions and randomization.

% Please add the following required packages to your document preamble:
% \usepackage{graphicx}
\begin{table}[]
\centering
\tiny
\begin{tabular}{M{2cm}|M{0.8cm}M{0.8cm}M{0.8cm}M{0.8cm}}
\hline
 & \multicolumn{2}{c}{\centering Non-Retrieval} & \multicolumn{2}{c}{\centering Retrieval} \\ \hline
Model & ASR & FPR & ASR & FPR \\ \hline
Vicuna-7B & 0.34 & 0.02 & 0.12 & 0.05 \\
Vicuna-13B & 0.27 & 0.01 & 0.07 & 0.03 \\
Vicuna-33B & 0.16 & 0.02 & 0.06 & 0.01 \\
Llama-7B & 0.22 & 0.01 & 0.06 & 0.01 \\
Llama-13B & 0.17 & 0.01 & 0.06 & 0.02 \\
Llama-70B & 0.14 & 0.01 & 0.05 & 0.01 \\ \hline
\end{tabular}%
\caption{We compare \textbackslash{}sysname with retrieval and non-retrieval. For retrieval-\sysname, the \sysname will perform the retrieval process to get the one-shot learning example for prompt decouple. While for non-retrieval-\sysname, the \textbackslash{}sysname directly performs prompt decouple.}
\label{tab:retrieval}
\end{table}

\subsection{Effect of Retrieval}\label{sec:eval:retrieval}
The Retrieval strategy, as indicated in the results (see Table. \ref{tab:retrieval}), plays a pivotal role in mitigating the risk of successful attacks (ASR) and in minimizing false alarms (FPR). When comparing the retrieval against non-retrieval settings, it's clear that the retrieval mechanism contributes to a reduction in both ASR and FPR for Llama-2 and Vicuna-1.5 models. Specifically, in non-retrieval scenarios, ASR for Llama-2 stands at 0.45 and 0.54 for Vicuna-1.5, which signifies a higher vulnerability to attacks when the system doesn't employ the retrieval method. Conversely, when retrieval is applied, there's a noticeable drop in ASR to 0.25 for Llama-2 and 0.31 for Vicuna-1.5, indicating a more robust defense posture. Furthermore, the FPR also shows a decline with retrieval, suggesting that the system becomes more accurate in distinguishing between benign and malicious queries, thus reducing the likelihood of legitimate queries being incorrectly flagged as attacks. Furthermore, the retrieval would not take much more time cost (see Fig. \ref{fig:time}).

\subsection{\rebuttal{The Un-retrieval Attack}}\label{sec:eval:unretrieval}
\begin{figure}[htbp]
  \centering
    \subfloat[\scriptsize{Llama}]{\includegraphics[width=0.24\textwidth]{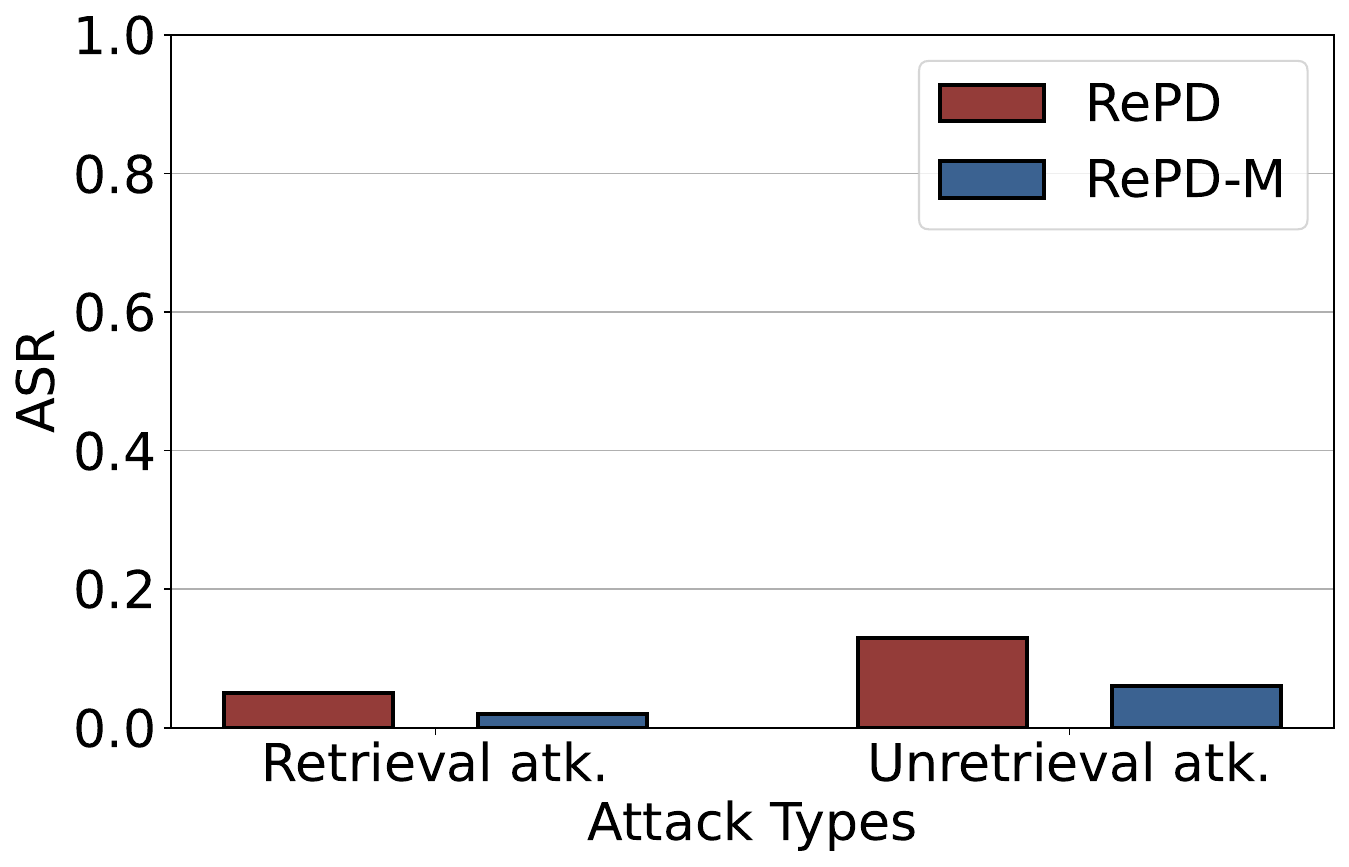}}
    \subfloat[\scriptsize{Vicuna}]{\includegraphics[width=0.24\textwidth]{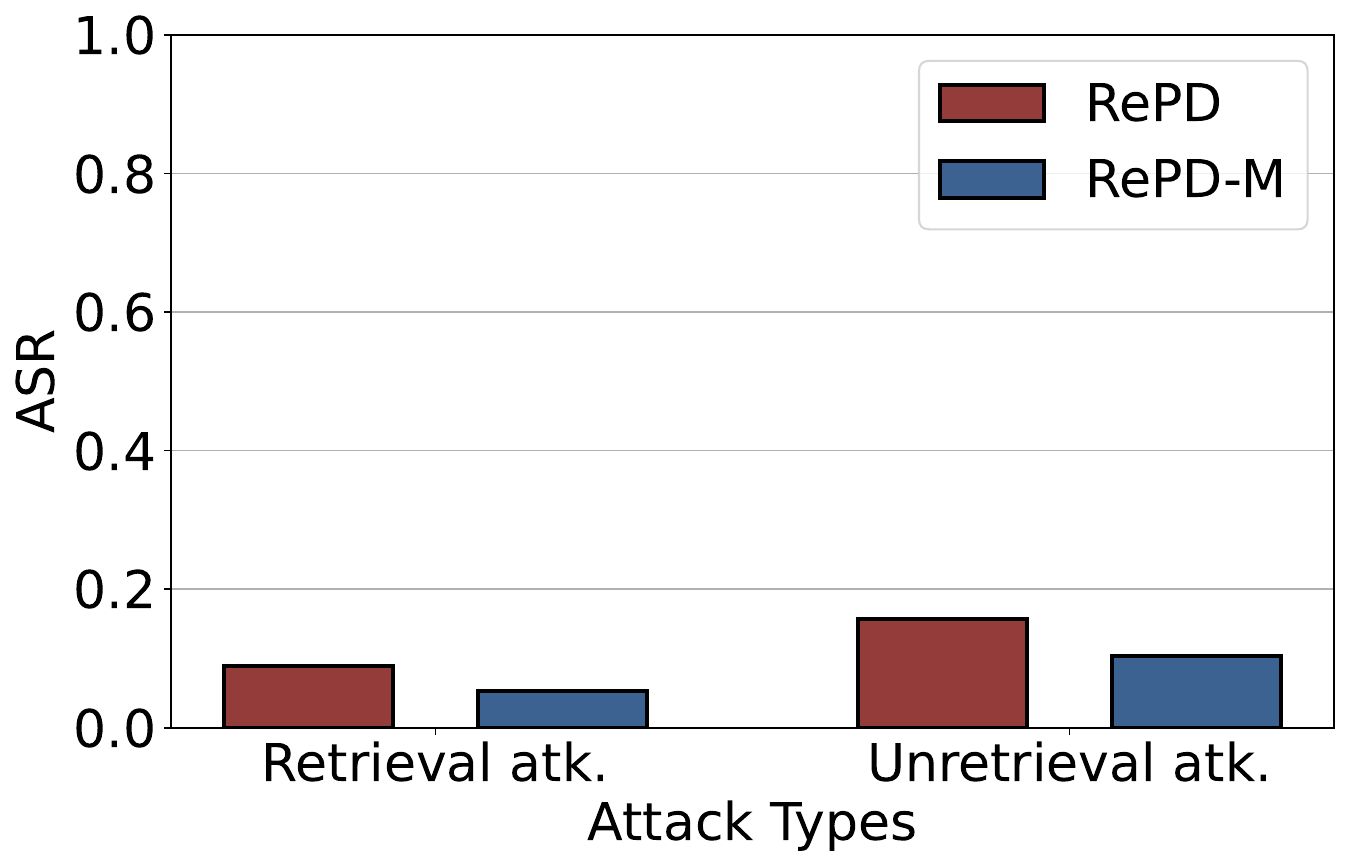}}
    \\
  \caption{
  \rebuttal{We compare \sysname's ability to handle the attacks stored in the retrieval database with those unstored in the database.}
  }
  \label{fig:unretrieval}
  \vspace{-10pt}
\end{figure}
\rebuttal{Considering \sysname needs to cope with the unseen attacks that are unstored in the retrieval database in the real-world settings, we compare \sysname's ability to handle the attacks stored in the retrieval database with those unstored in the database (see Figure. \ref{fig:unretrieval}).
The evaluation results indicate that, though the ASR on un-retrieval attacks increases a little compared with the retrieval attacks, the absolute value of it still remains under 0.15.
The rationale for the defense effectiveness is that the problem decouple process itself can defend the jailbreak attacks already (also evaluated in \S\ref{sec:eval:retrieval}).
While the retrieval process and the problem decouple as a one-shot learning example provides a better defense against the retrieved ones.}
\section{Limitation}

\par Though \sysname can achieve better defense performance than previous methods 
%against template-based jailbreak attacks
, the approach still introduces extra time costs due to extending token length. 
Furthermore,
% due to our template-based attack definition, 
the main goal of \sysname is to defend against templated-based attacks. 
We leave the defense method against potential future emerging attacks, which are out of the scope of templated-based jailbreak as future work. 
% Furthermore, the response quality will be impacted as well since the query prompt is edited through this process.

\section{Conclusion}

\par In conclusion, this paper introduces \sysname, a novel defense framework designed to counteract jailbreak attacks on large language models (LLMs). Despite the extensive pre-training and fine-tuning in moral alignment, LLMs are still susceptible to generating harmful information when prompted by users. \sysname addresses this vulnerability by employing an attack-retrieval-based prompt decomposition strategy. This framework leverages a retrieval database to construct a one-shot learning example, enabling the LLM to decompose tasks from prompts by recognizing and mitigating known attacks.

\par Our experimental validation
%conducted across a broad spectrum of harmful and benign prompts, 
demonstrates the efficacy of \sysname in assisting LLMs against jailbreak attacks. The evaluation results prove that \sysname will not impact the benign response.

\newpage
\bibliography{custom}

\appendix
\newpage

\section{Appendix}

\subsection{Settings for Adaptive Attacks}\label{sec:appendix:adaptivate}

\par In this section, we introduce the setting of the adaptive attack in the main paper.

\par For GCG, AutoDAN, PAIR, and TAP, their authors have released the code for reproduction. When using their codes, we leave most of the default settings unchanged. The setting is set as:

\begin{itemize}
    \item GCG: n-steps=1000, test-steps=50, batch-size=512
    \item AutoDAN:  num-steps=100, batch-size=64
    \item PAIR:  n-stream=5, n-iterations=12
    \item TAP: n-stream=1, branching-factor=2, width=5, depth=5
\end{itemize}

\subsection{The Prompt for \sysname}
\begin{prompt}
\begin{mdframed}[linecolor=black,linewidth=1pt]
    \sssec{\sysname's prompt with retrieval.}
    \par For the prompt that the user enters \textcolor{red}{$\$user\_prompt$}, you need to decouple it into two parts first: the template of the user's query method and the user's question or request. I will give you an example: for example, the user asks: \textcolor{blue}{$\$full\_prompt$}, you need to decouple it into query template: \textcolor{blue}{$\$prompt\_template$} and the user's question or request: \textcolor{blue}{$\$prompt\_request$}. Then, in your response, you need to state the request or the question by the user again, then state your official response. Remember, refuse to generate harmful responses.
\end{mdframed}
\label{prompt:retrievalPrompt}
\end{prompt}
\par The above Prompt. \ref{prompt:retrievalPrompt} is the retrieval prompt template that \sysname adopts. Within the prompt template, \textcolor{blue}{$\$full\_prompt$} is the full prompt which is constructed by \textcolor{blue}{$\$prompt\_template$} and \textcolor{blue}{$\$prompt\_request$}. \textcolor{blue}{$\$prompt\_template$} is the retrieval jailbreak template, while \textcolor{blue}{$\$prompt\_request$} is the randomly selected harmful request. \textcolor{red}{$\$user\_prompt$} denotes the prompt query user inputs.

\begin{prompt}
\begin{mdframed}[linecolor=black,linewidth=1pt]
    \sssec{\sysname's prompt without retrieval.}
    \par For the prompt query that the user enters \textcolor{red}{$\$user\_prompt$}, you need to decouple the prompt's real question, which is the question you want to answer. Remember, to state the question in your response first, then continue your official response to the query. Remember, refuse to generate harmful responses.
\end{mdframed}
\label{prompt:nonRetrievalPrompt}
\end{prompt}
\par The above Prompt. \ref{prompt:nonRetrievalPrompt} is the non-retrieval prompt template which only decouples the harmful question but without a retrieval process.

\begin{figure}[htbp]
  \centering
    \subfloat[\scriptsize{Vicuna-ASR}]{\includegraphics[width=0.16\textwidth]{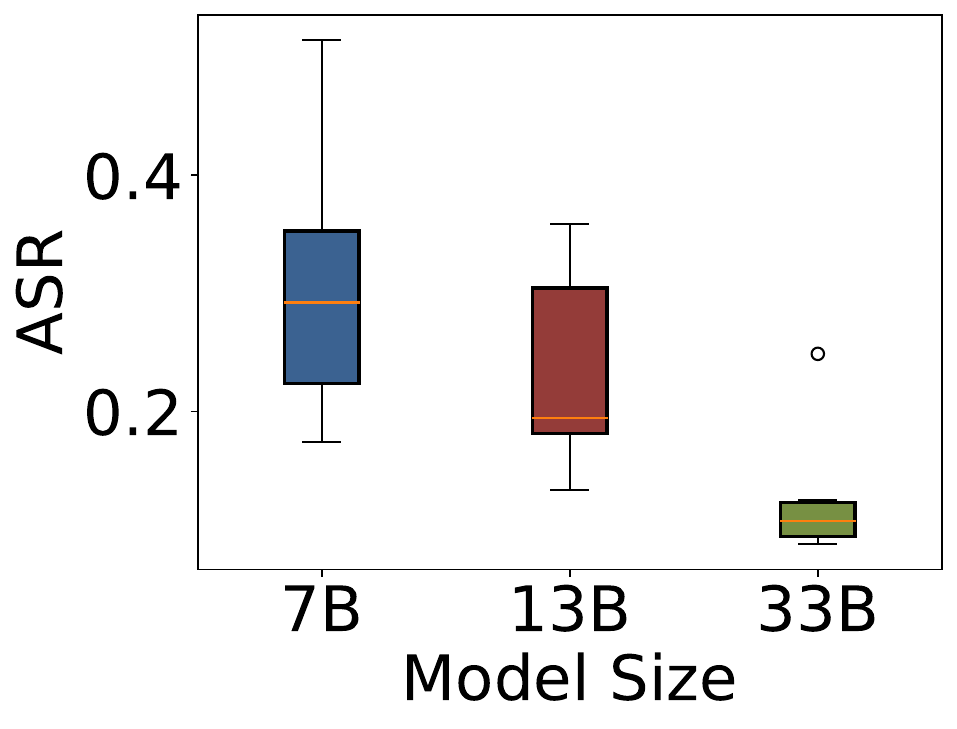}}
    \subfloat[\scriptsize{Vicuna-FPR}]{\includegraphics[width=0.16\textwidth]{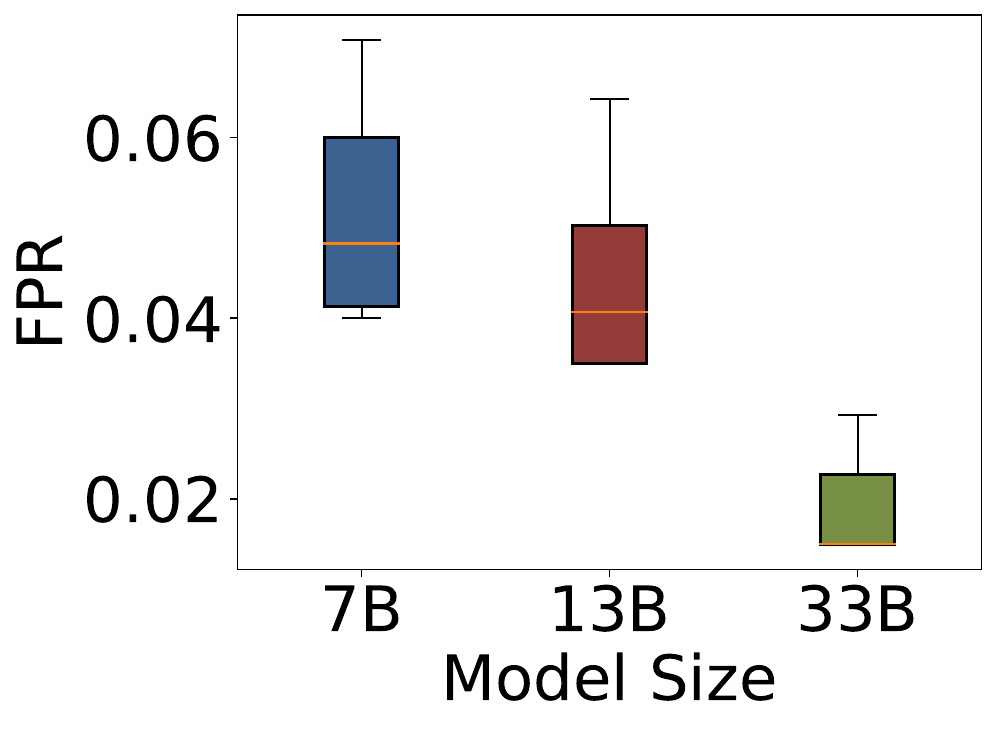}}
    \subfloat[\scriptsize{Vicuna-accuracy}]{\includegraphics[width=0.16\textwidth]{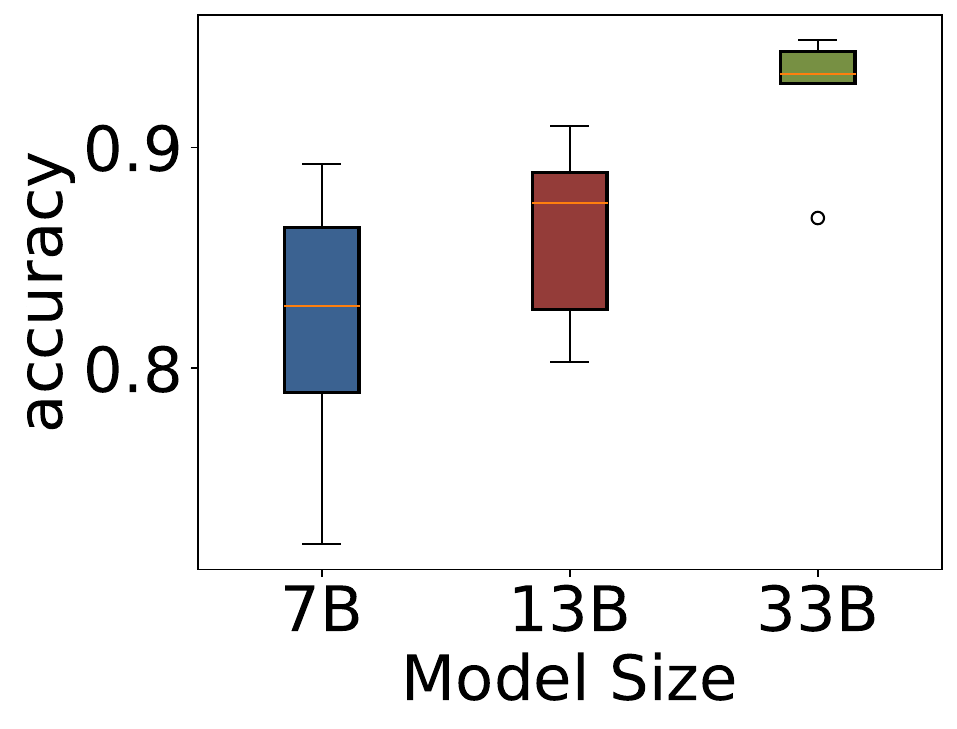}}
    \\
    \subfloat[\scriptsize{Llama-ASR}]{\includegraphics[width=0.16\textwidth]{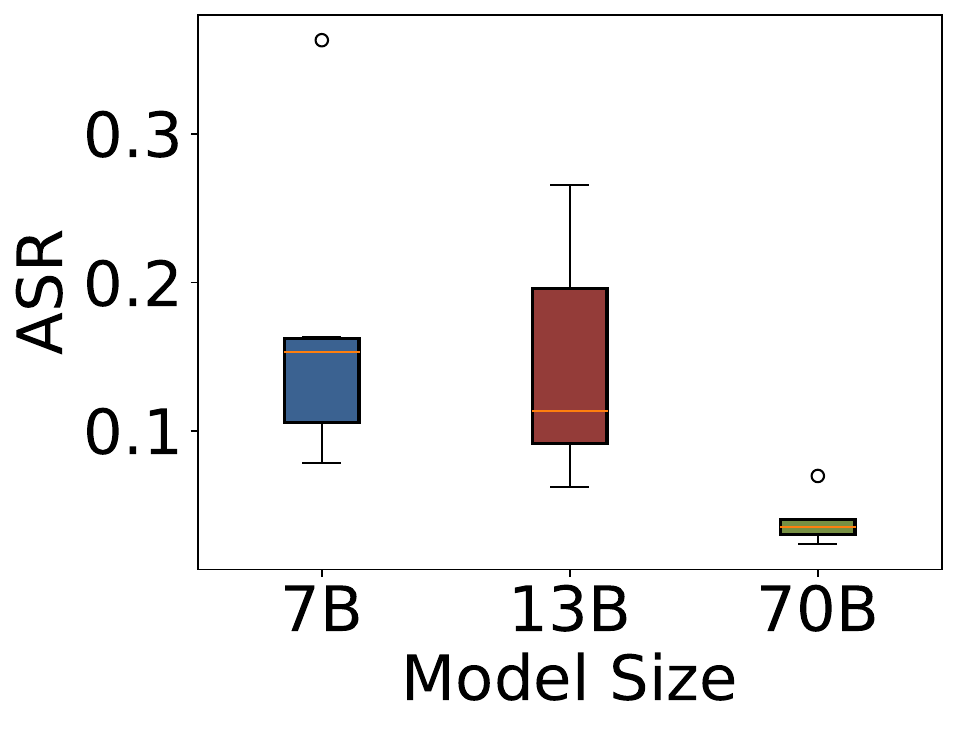}}
    \subfloat[\scriptsize{Llama-FPR}]{\includegraphics[width=0.16\textwidth]{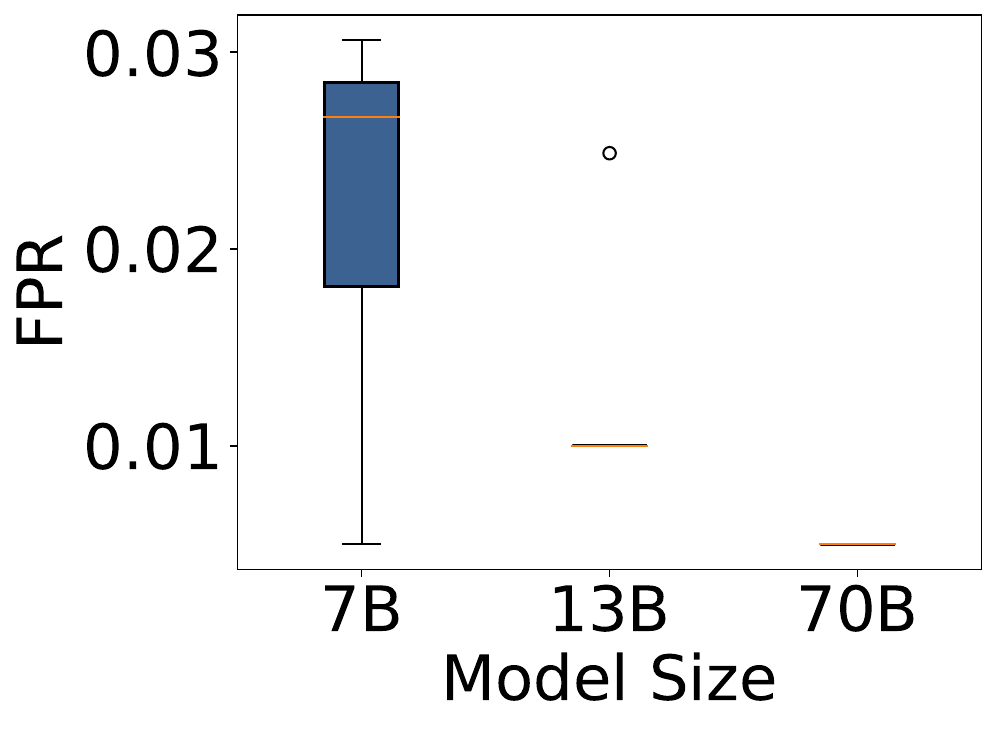}}
    \subfloat[\scriptsize{Llama-accuracy}]{\includegraphics[width=0.16\textwidth]{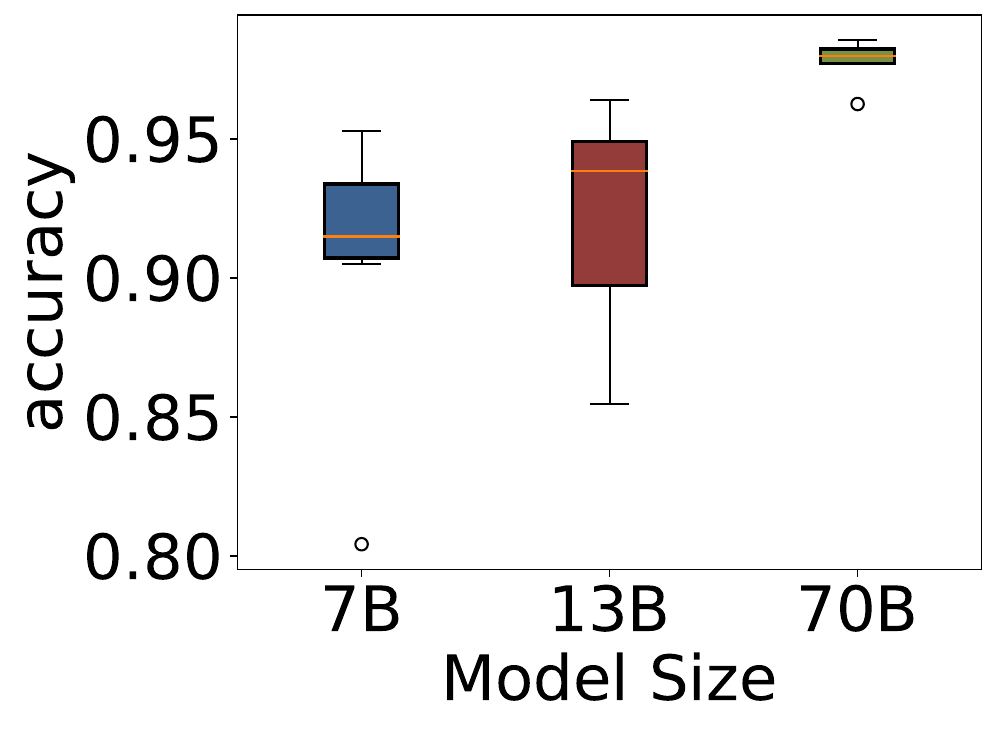}}
    \\
  \caption{
  Evaluation of \sysname's performance on different model sizes.
  }
  \label{fig:model_size}
  \vspace{0.2in}
\end{figure}

\subsection{Benign Question Generation Process for \sysname's Evaluation}
\begin{prompt}
\begin{mdframed}[linecolor=black,linewidth=1pt]
    Generate a list of questions. These questions should be one sentence long and have a clear goal and intention. The intention of the generated questions must be benign and legal. The questions should cover various fields, including life, geography, history, skills, etc.
\end{mdframed}
\label{prompt:questionGenerate}
\end{prompt}

\par We used the prompt within Prompt. \ref{prompt:questionGenerate} to input into ChatGPT-4 to generate 200 benign questions for our evaluation.

%\subsection{Retrieval Database Construction}

\end{document}